\documentclass[aps,prx,amsmath, amssymb,reprint,superscriptaddress]{revtex4-2}

\usepackage{graphicx}

\usepackage{xcolor}

\definecolor{burgundy}{rgb}{0.5, 0.0, 0.13}
\definecolor{bole}{rgb}{0.47, 0.27, 0.23}

\usepackage[range-phrase=\text{--}, range-units=single]{siunitx}
\newcommand{\fref}[1]{Figure \ref{#1}}
\newcommand{\fs}{\femto\second}

\begin{document}


\title{Correlation Driven Transient Hole Dynamics Resolved in Space and Time in the Isopropanol Molecule}



\author{T.~Barillot}
\affiliation{Quantum Optics and Laser Science Group, Blackett Laboratory, Imperial College London, London, SW7~2BW, UK.}

\author{O.~Alexander}
\affiliation{Quantum Optics and Laser Science Group, Blackett Laboratory, Imperial College London, London, SW7~2BW, UK.}

\author{B.~Cooper}
\affiliation{Quantum Optics and Laser Science Group, Blackett Laboratory, Imperial College London, London, SW7~2BW, UK.}
\affiliation{Atomic, Molecular, Optical and Positron Physics Group, Department of Physics and Astronomy, University College London, Gower Street
London, WC1E 6BT, UK.}

\author{T.~Driver}
\thanks{Corresponding author. tdriver@stanford.edu}
\affiliation{Quantum Optics and Laser Science Group, Blackett Laboratory, Imperial College London, London, SW7~2BW, UK.}
\affiliation{Stanford PULSE Institute, SLAC National Accelerator Laboratory , USA}
\affiliation{Linac Coherent Light Source, SLAC National Accelerator Laboratory, Menlo Park, California, 94025, USA}

\author{D.~Garratt}
\affiliation{Quantum Optics and Laser Science Group, Blackett Laboratory, Imperial College London, London, SW7~2BW, UK.}

\author{S.~Li}
\affiliation{Accelerator Research Division, SLAC National Accelerator Laboratory, Menlo Park, California, 94025, USA}

\author{A.~Al~Haddad}
\affiliation{Chemical Sciences and Engineering Division, Argonne National Laboratory, Argonne, IL 60439, United States of America}
\affiliation{Paul-Scherrer Institute, CH-5232 Villigen PSI, Switzerland}

\author{A.~Sanchez-Gonzalez}
\affiliation{Quantum Optics and Laser Science Group, Blackett Laboratory, Imperial College London, London, SW7~2BW, UK.}

\author{M.~Ag\aa ker}
\affiliation{Department of Physics and Astronomy, Uppsala University, Box 516, 751 20, Uppsala, Sweden}
\affiliation{MAX IV Laboratory, Sweden}

\author{C.~Arrell}
\affiliation{Paul-Scherrer Institute, CH-5232 Villigen PSI, Switzerland}

\author{M.~Bearpark}
\affiliation{Department of Chemistry, Imperial College London, London SW7 2AZ, United Kingdom}

\author{N.~Berrah}
\affiliation{Department of Physics, University of Connecticut, USA}

\author{C.~Bostedt}
\affiliation{Chemical Sciences and Engineering Division, Argonne National Laboratory, Argonne, IL 60439, United States of America}
\affiliation{Paul-Scherrer Institute, CH-5232 Villigen PSI, Switzerland}
\affiliation{LUXS Laboratory for Ultrafast X-ray Sciences, Institute of Chemical Sciences and Engineering, \'Ecole Polytechnique F\'ed\'erale de Lausanne (EPFL), CH-1015 Lausanne, Switzerland}

\author{J.~Bozek}
\affiliation{Synchrotron SOLEIL, France}

\author{C.~Brahms}
\affiliation{Quantum Optics and Laser Science Group, Blackett Laboratory, Imperial College London, London, SW7~2BW, UK.}

\author{P.~H.~Bucksbaum}
\affiliation{Stanford PULSE Institute, SLAC National Accelerator Laboratory , USA}

\author{A.~Clark}
\affiliation{Laboratory of Molecular Nanodynamics, \'Ecole Polytechnique F\'ed\'erale de Lausanne, CH-1015 Lausanne, Switzerland}

\author{G.~Doumy}
\affiliation{Chemical Sciences and Engineering Division, Argonne National Laboratory, Argonne, IL 60439, United States of America}

\author{R.~Feifel}
\affiliation{Department of Physics, University of Gothenburg, Origivaegen 6 B, 412 58, Gothenburg, Sweden}

\author{L.J.~Frasinski}
\affiliation{Quantum Optics and Laser Science Group, Blackett Laboratory, Imperial College London, London, SW7~2BW, UK.}

\author{S.~Jarosch}
\affiliation{Quantum Optics and Laser Science Group, Blackett Laboratory, Imperial College London, London, SW7~2BW, UK.}

\author{A.~S.~Johnson}
\affiliation{Quantum Optics and Laser Science Group, Blackett Laboratory, Imperial College London, London, SW7~2BW, UK.}

\author{L.~Kjellsson}
\affiliation{Department of Physics and Astronomy, Uppsala University, Box 516, 751 20, Uppsala, Sweden}

\author{P.~Koloren\v{c}}
\affiliation{Charles University, Faculty of Mathematics and Physics, Institute of Theoretical Physics, V Hole\v{s}ovi\v{c}k\'{a}ch 2, 180 00 Prague, Czech Republic.
}

\author{Y.~Kumagai}
\affiliation{Chemical Sciences and Engineering Division, Argonne National Laboratory, Argonne, IL 60439, United States of America} 

\author{E.~W.~Larsen}
\affiliation{Quantum Optics and Laser Science Group, Blackett Laboratory, Imperial College London, London, SW7~2BW, UK.}

\author{P.~Matia-Hernando}
\affiliation{Quantum Optics and Laser Science Group, Blackett Laboratory, Imperial College London, London, SW7~2BW, UK.}

\author{M.~Robb}
\affiliation{Department of Chemistry, Imperial College London, London SW7 2AZ, United Kingdom}

\author{J.-E.~Rubensson}
\affiliation{Department of Physics and Astronomy, Uppsala University, Box 516, 751 20, Uppsala, Sweden}

\author{M.~Ruberti}
\affiliation{Quantum Optics and Laser Science Group, Blackett Laboratory, Imperial College London, London, SW7~2BW, UK.}

\author{C.~Sathe}
\affiliation{MAX IV Laboratory, Sweden}

\author{R.~J.~Squibb}
\affiliation{Department of Physics, University of Gothenburg, Origivaegen 6 B, 412 58, Gothenburg, Sweden}

\author{A.~Tan} 
\affiliation{Quantum Optics and Laser Science Group, Blackett Laboratory, Imperial College London, London, SW7~2BW, UK.}

\author{J.~W.~G.~Tisch} 
\affiliation{Quantum Optics and Laser Science Group, Blackett Laboratory, Imperial College London, London, SW7~2BW, UK.}

\author{M.~Vacher} 
\affiliation{Department of Chemistry, Imperial College London, London SW7 2AZ, United Kingdom}
\affiliation{Université de Nantes, CNRS, CEISAM UMR 6230, F-44000 Nantes, France.}

\author{D.~J.~Walke}
\affiliation{Quantum Optics and Laser Science Group, Blackett Laboratory, Imperial College London, London, SW7~2BW, UK.}

\author{T.~J.~A.~Wolf}
\affiliation{Stanford PULSE Institute, SLAC National Accelerator Laboratory , USA}

\author{D.~Wood}
\affiliation{Quantum Optics and Laser Science Group, Blackett Laboratory, Imperial College London, London, SW7~2BW, UK.}

\author{V.~Zhaunerchyk}
\affiliation{Department of Physics, University of Gothenburg, Origivaegen 6 B, 412 58, Gothenburg, Sweden}

\author{P.~Walter}
\affiliation{Linac Coherent Light Source, SLAC National Accelerator Laboratory, Menlo Park, California, 94025, USA}
\author{T.~Osipov}
\affiliation{Linac Coherent Light Source, SLAC National Accelerator Laboratory, Menlo Park, California, 94025, USA}
\author{A.~Marinelli}
\affiliation{Accelerator Research Division, SLAC National Accelerator Laboratory, Menlo Park, California, 94025, USA}
\author{T.~Maxwell} 
\affiliation{Accelerator Research Division, SLAC National Accelerator Laboratory, Menlo Park, California, 94025, USA}
\author{R.~Coffee}
\affiliation{Linac Coherent Light Source, SLAC National Accelerator Laboratory, Menlo Park, California, 94025, USA}
\author{A.~A.~Lutman}
\affiliation{Accelerator Research Division, SLAC National Accelerator Laboratory, Menlo Park, California, 94025, USA}
\author{V.~Averbukh}
\affiliation{Quantum Optics and Laser Science Group, Blackett Laboratory, Imperial College London, London, SW7~2BW, UK.}
\author{K.~Ueda}
\affiliation{Institute of Multidisciplinary Research for Advanced Materials, Tohoku University, Sendai 980-8577, Japan }
\author{J.~P.~Cryan}
\affiliation{Stanford PULSE Institute, SLAC National Accelerator Laboratory , USA}
\affiliation{Linac Coherent Light Source, SLAC National Accelerator Laboratory, Menlo Park, California, 94025, USA}
\author{J.~P.~Marangos}
\affiliation{Quantum Optics and Laser Science Group, Blackett Laboratory, Imperial College London, London, SW7~2BW, UK.}


\date{\today}

\begin{abstract}
The possibility of suddenly ionized molecules undergoing extremely fast electron hole dynamics prior to significant structural change was first recognized more than 20 years ago and termed~\textit{charge migration}.  
The accurate probing of ultrafast electron hole dynamics requires measurements that have both sufficient temporal resolution and can detect the localization of a specific hole within the molecule. We report an investigation of the dynamics of inner valence hole states in isopropanol where we use an x-ray pump/x-ray probe experiment, with site and state-specific probing of a transient hole state localized near the oxygen atom in the molecule, together with an \textit{ab initio} theoretical treatment. 
We record the signature of transient hole dynamics and make the first observation of dynamics driven by frustrated Auger-Meitner transitions. We verify that the hole lifetime is consistent with our theoretical prediction. 
This state-specific measurement paves the way to widespread application for observations of transient hole dynamics localized in space and time in molecules and thus to charge transfer phenomena that are fundamental in chemical and material physics.

\end{abstract}


\maketitle

\section{Introduction}
One of the most fundamental problems in photo-physics is the electronic response of a quantum system to an impulse. 
Impulsive excitation creates time-dependent oscillations in the charge density, i.e. an electronic wavepacket or superposition of electronic states. 
This electronic superposition resulting from ionization produces rapid evolution in the electronic hole density that may result in movement of charge density within the molecular frame, a process termed ``charge migration'' ~\cite{cederbaum1999ultrafast}. 

While this initial charge dynamics is purely electronic in nature, eventually this charge disturbance will couple to other degrees of freedom in the system such as nuclear motion, leading to a localization of the charge. 
This transfer of electronic charge across molecular bonds is fundamental to an understanding of charge transfer phenomena~\cite{may2008charge,worner2017charge}, essential in the quantum description of various biological and electrochemical processes. 
A deeper knowledge of these fundamental processes, including concepts of charge directed reactivity~\cite{remacle_charge_1998},  may be used to steer photo-chemical reactions.

Recent technological developments in generating short wavelength radiation has enabled laser-driven impulses in gas phase and condensed phase systems~\cite{krausz_attosecond_2009,goulielmakis_real-time_2010,cavalieri_attosecond_2007,neppl_attosecond_2012}.
These short wavelength laser pulses can efficiently drive removal of inner valence electrons.
For cations of moderately sized molecules, where the Auger-Meitner decay~\cite{Auger1923, Meitner1922, Meitner1923} of the inner-shell vacancy is energetically forbidden, electron correlation connects the initial inner valence vacancy state~(or hole state) of the molecule to a discrete set of eigenstates of the cation spaced by \SIrange{0.1}{1}{\eV}. 
This is because these high-lying cationic eigenstates are not well described by a single electron configuration~\cite{kuleff_ultrafast_2014}. 

\begin{figure}
    \centering
    \includegraphics[width=0.49\textwidth]{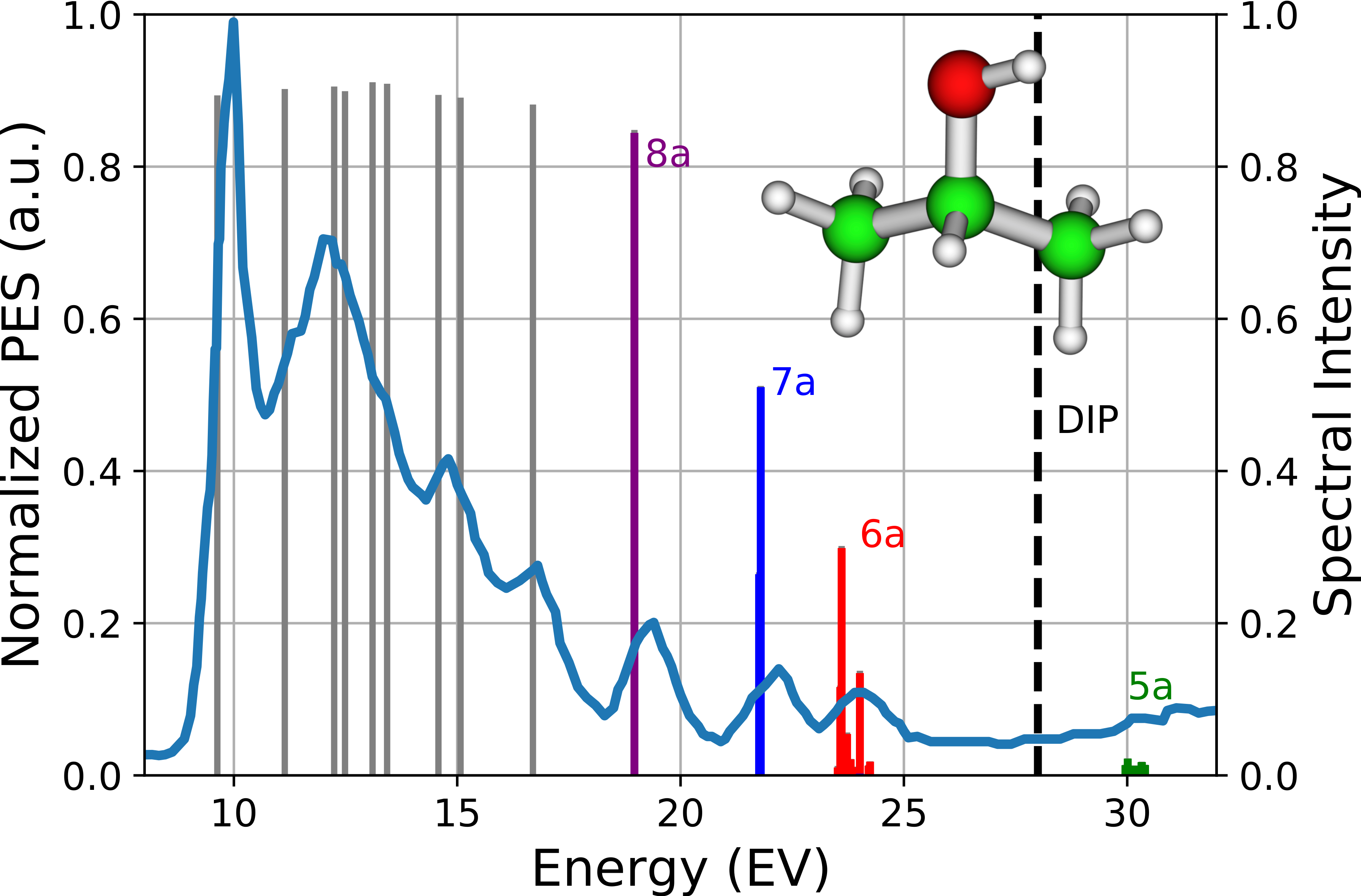}
    \caption{Photoelectron spectrum of isopropanol, recorded using XUV synchrotron radiation at 90~eV (blue). Gray vertical lines mark the spectral intensity calculated using ADC(2)x, colored lines are contributions to the spectral intensity of different orbitals 8a~(purple) to 5a~(green). Black dashed line denotes the double ionization potential of isopropanol at 28~eV, measured at EPFL using spectrally filtered HHG in a photoelectron-photoion coincidence (PEPICO) set up. Inset: Ball and stick diagram of the isopropanol gauche conformer. Our resonant x-ray probe interrogates the local electron density at the oxygen site (red).}
    \label{fig:Isopropanol_PES}
\end{figure}

In our work we implement a new experimental method to study the dynamics of inner valence hole states, which provides state- and site-specificity along with the requisite temporal resolution. 
We report new results that advance our understanding of decoherence and charge transfer phenomena following impulsive ionization. 
We study inner valence hole~(IVH) states in the isopropanol cation. 
Distinct types of temporal evolution are found for superpositions of cation states originating from the removal of an electron from different molecular orbitals.   
In \fref{fig:Isopropanol_PES}
we show the photoelectron spectrum of the isopropanol valence states (recorded using XUV synchrotron radiation at $90$~eV). 
These cationic states are labeled according to the position of the hole in the dominant electron configuration, with the inner valence states $7a^{-1}$ and $6a^{-1}$ being of primary interest in the current study. 
We probe the electron-hole states following impulsive x-ray ionization using time-resolved measurements supported by \textit{ab initio} theory. 

 Many-electron theory distinguishes two types of charge migration: (a) where there are a small number of states with different spatial locations within the molecule so that the oscillating charge cloud moves back and forth across the molecule~\cite{kuleff2014ultrafast} and (b) where one localized hole state is coupled to a large series of de-localized states so the charge motion is of the form of a ``breathing" of the charge density around a single center ~\cite{cederbaum1999ultrafast,AverbukhPRL2010}. 
 The paradigm for the case (a) is the familiar double-well system where population of a gerade/ungerade state superposition results in the probability density oscillations between the two centers. 
 The paradigm for the case (b) is fundamentally different - it is the Bixon-Jortner model for coupling of a bound state to a quasi-continuum ~\cite{BixonJortnerJCP1967,CooperAverbukhJCP2014} predicting initial exponential decay followed by partial revivals. 
 While the case~(a) has been observed indirectly~(i.e. not via electronic observables) in experiments~\cite{calegari2014ultrafast,worner2017charge}, case~(b) has so far proven to be more elusive for  experimental detection. 
 Here, we report a direct observation of case~(b), namely of the  electronic motion that results in the quasi-exponential decay of the hole density in the vicinity of the initial localization. 
 The electron hole formed in our experiment interacts strongly with a series of Auger-Meitner type states, but has insufficient energy to decay through the emission of an Auger-Meitner electron. The mechanism is often referred to as frustrated Auger-Meitner configurations~\cite{Bagus2004}. 
 As the electron density cannot escape the molecular volume there is some probability of return to its initial state before nuclear motion-driven dephasing occurs. 
 Whether or not that return results in an observable oscillation in the hole density depends upon the degree of coherence of the superposition.

In principle, the states of the cation should be described by the full wavefunction, including both electronic and nuclear components. 
In practice, for polyatomic molecules it is often necessary to look to approximations for the inclusion of the nuclear part of the wavefunction. 
The current state of the art can allow coupling of some tens of nuclear degrees of freedom to some few electronic excited states~\cite{Kuleff, Marciniak}, which has been applied to the study of ionization from HOMO orbitals in propiolic acid\cite{Kuleff}  as well as deeper bound states in naphthalene\cite{Marciniak}, exhibiting frustrated auto-ionizing states.

Here, we treat the electronic interactions at a sufficiently high accuracy (via the ADC(2)x method) to include the electronic correlations in the system. This is especially important for the inner valence states we are interested in, where charge migration is predicted to accompany frustrated Auger-Meitner transitions. 
To understand the importance of nuclear degrees of freedom in the charge dynamics, we conceptually separate the dynamical considerations into two parts, (i) the nuclear zero point geometry spread (convolved with the inherent geometry spread due to a finite temperature sample) that introduces a fast dephasing to any coherent electronic dynamics ~\cite{vacher2017electron}, (ii) and the nuclear motion of the cation subsequent to ionization. 

It is understood that there is strong influence from coupling to the nuclear part of the wavefunction through the zero-point geometry spread. This zero-point geometry spread will cause dephasing of the electronic superposition and so influences the observation of electronic state evolution even on a few femtosecond timescale.  Simulations including the zero point spread in nuclear geometry indicate, see Section~\ref{sec:theory}, that although decoherence of the electronic wavepacket reduces visibility of any oscillatory dynamics the initial fast decay should still be observable. 
On timescales typically greater than \SI{10}{\fs}, but with evidence that these effects can manifest faster than this ~\cite{Baker,HJW,Johnson,Mikosch}, the dynamics of the different nuclear modes excited by the impulse also become important ~\cite{vacher2017electron,vacher2015electronPRA,mendive2013coupled}.
Thus, when considering an ensemble of molecules, it becomes essential to consider the distribution of molecules within the nuclear phase space (both position and momentum) at the time of the impulse and then to follow the dynamics, demanding the most advanced experimental and theoretical (i.e. with classical nuclear dynamics~\cite{worth2008solving,vacher2015electron} and fully quantum approaches~\cite{jenkins2018ehrenfest}) tools available to further our understanding.

Our measurement realizes a new method for probing the dynamical evolution of the electronic wavepacket produced by impulsive ionization. 
Our calculations show that the dynamics of the 6$a$ hole states are indeed dominated by frustrated Auger-Meitner processes, which leads to rapid decay of the localized state, but crucially without the energetically forbidden secondary electron emission.
Since the emission of a secondary Auger-Meitner electron is energetically forbidden we cannot use measurements that rely on the direct Auger-Meitner emission from these states such as photoelectron streaking~\cite{drescher_time-resolved_2002,haynes_clocking_2020}.
Instead we must use strategies to track the hole state evolution by probing directly the time-dependent hole state probability. 
To this end we monitor the delay dependent amplitude of resonant oxygen $1s$ to inner valence excitations in the cation using a transient absorption methodology with short (few-fs) x-ray pulse.

Earlier time-resolved measurements of impulse driven charge motion relied on an XUV-pump/strong infrared~(IR) probe scheme~\cite{calegari2014ultrafast,perfetto2020ultrafast}, where the pump pulse is derived from a high harmonic generation~(HHG) based source.
Despite a clear few femtosecond response of the system, a complete description of the influence of the IR field on the system dynamics is challenging within current theoretical capabilities, which makes a direct comparison with theory difficult. 

In our work we show that by using advanced ultrashort x-ray pulses we can unambiguously monitor the inner valence hole decay with a state-specific, atomically localized probe that does not perturb the dynamics.
With the advent of x-ray free electron lasers (XFEL), few femtosecond~\cite{lutman2016fresh,lutman2018high} and more recently attosecond~\cite{duris2020tunable} x-ray pulses are available with sufficient energy to both initiate and probe the cation dynamics via single photon processes. 
In addition to simplifying the interpretation of the measurement~\cite{lara2016decoherence}, state and atomic site specificity can be attained by varying the probe photon energy.
Our results demonstrate the first observation of states with dynamics driven by frustrated Auger-Meitner transitions and pave the way to quantitative studies of the onset and decay of quantum coherence in larger molecules and condensed phase systems.

\section{X-ray pump-probe measurement for site specific, field free observation}

\begin{figure*}
    \centering
    \includegraphics{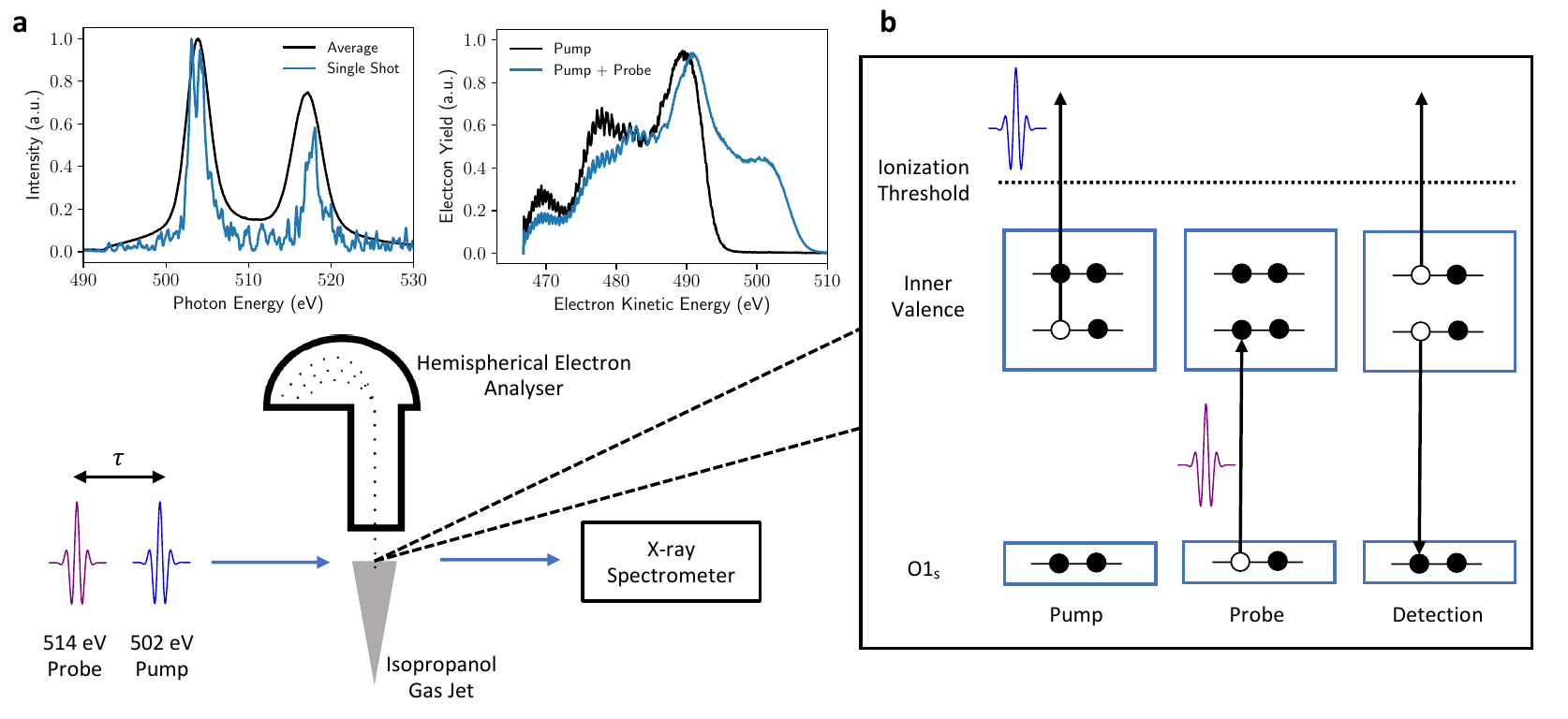}
    \caption{(a) Experimental set-up with two colour pulses (pump central photon \SI{502}{\eV}, probe central photon energy \SI{514}{\eV}) from the fresh-slice mode with delay controlled by chicane, the interaction point equipped with a hemispherical electron analyzer and a down-stream x-ray spectrometer to determine the pulse spectra and relative energies on every shot. (b) Pump step opens a number of valence ionization channels including creation of inner valence holes (IVH). The delayed probe pulse can strongly interact via a O 1$a$–IVH transition (a channel only open if inner valence ionization has occurred). Following this there can be Auger-Meitner decay back to the O 1$a$ hole with the emission of Auger-Meitner electrons of characteristic energy that can be detected.}
    \label{fig:ExperimentalLayout}
\end{figure*}

We measure gas phase isopropanol using an x-ray pump - x-ray probe pulse scheme at the AMO endstation of the Linac Coherent Light Source~(LCLS) free electron laser~\cite{amo1, amo2}. 
At each shot we record both the kinetic energy spectrum of the high-energy electrons produced by the x-ray/matter interaction and the photon spectrum of the two x-ray pulses.
Making use of the so-called ``fresh slice'' mode to generate two-color pairs of ultra-short x-ray pulses with a variable inter-pulse delay~\cite{lutman2016fresh}. 
In a first set of measurements the x-ray pulses were $\sim$7~fs in duration, and in a second set of measurements they were reduced to around 2.5~fs in duration.
In either case, a non-resonant pump pulse, with a photon energy of  $\sim$\SI{502}{eV}, is used to ionize the molecule.
Since the pulse has a coherent bandwidth of $\sim1.1$~eV, this ionization can form coherent superpositions of cationic states. 
We focus on superposition states created by electron correlation satellites of the inner valence states, as described above. 
However, it would also be possible to use this technique study superposition states that are composed of cationic states with a single outer-valence hole configuration.  

We target the wavepacket dynamics initiated by the removal of an electron from either the 6$a$ or 7$a$ molecular orbital~(see Figure 1 and discussion in Section III) and probe the hole formed by photoionization using a second x-ray pulse.
The probe pulse is tuned to be resonant with the transition from the $1s$ orbital located on the oxygen atom to the vacated 6$a$ and 7$a$ orbitals, which have spectrally isolated transition energies around \SI{514}{\eV} and \SI{516}{\eV} respectively (see \fref{fig:ExperimentalLayout}). 
Site specificity is achieved because the resonant dipole transitions between the oxygen $1s$ and IVH state depend critically on the hole density in the vicinity of the oxygen site. 
A similar scheme, envisioned for an equivalent IVH in glycine, was discussed in detail in a recent paper by Cooper \textit{et al.} \cite{cooper2014analysis}. 
Here, the probe is based upon the delay-dependent absorption of the resonant pulse into the vacant IVH. 
This leaves the molecule with an oxygen $1s$ core excitation, which will undergo rapid Auger-Meitner decay.
We measure the yield of Auger-Meitner electrons as a function of the incident x-ray photon energy and inter-pulse delay to determine the transient absorption spectrum~\cite{wolf2017probing}.  

This technique provides both site- and state-selectivity, which is critical since the pump pulse can produce any of the valence hole and carbon core hole states~(as is seen in Figure 1 and verified by calculation for 500 eV ionization pulse).
Moreover, there is a propensity for the ionization to produce outer valence hole states.
However, the dynamics of the outer valence states will occur in a different spectral region compared to the inner valence hole states, 
and thus the two signals can be easily distinguished using our x-ray probe.
Whilst in this work we study the dynamics following inner valence ionization, our method can be straightforwardly extended to the outer valence region by increasing the probe photon energy.
It is also important to note, that while the pump pulse can also remove electrons from any of the valence and carbon core orbitals, the energy ($502$~eV) is below threshold for oxygen $1s$ core ionization, so there is no signal in the Auger channels due to the pump alone.

\begin{figure}
    \centering
    \includegraphics[width=0.4\textwidth]{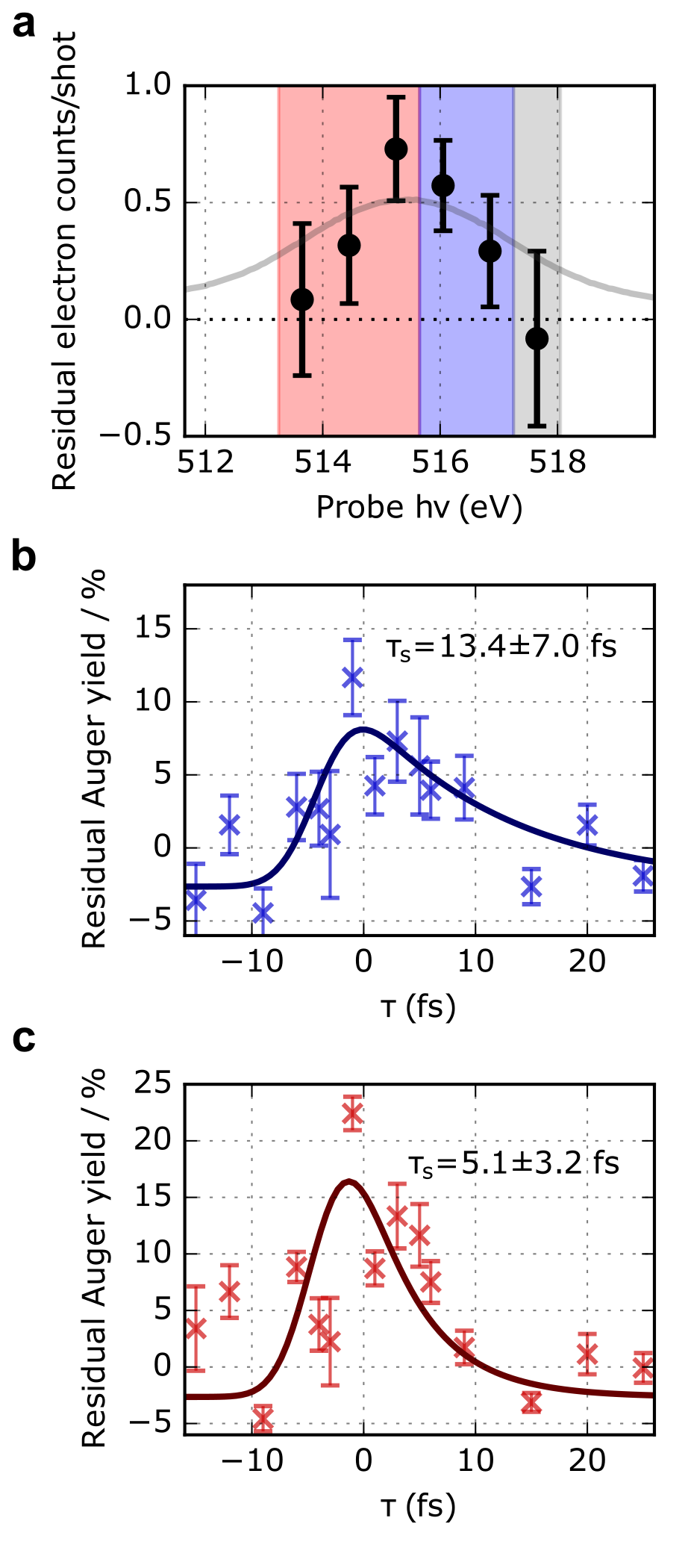} 
    \caption{\textbf{a} Probe central photon energy dependence of residual signal at delay $\sim$0~fs.\textbf{b} Time dependence of residual resonant high kinetic energy electron signal at higher probe photon energy, about expected 1$a$ $\rightarrow$ 7$a$ transition (blue), and \textbf{c} time dependence of residual resonant high energy electron yield  at lower probe photon energy about expected 1$a$ $\rightarrow$ 6$a$ transition (red). Fits to an exponentially modified Gaussian functional form are shown in black and the extracted decay time is given in each panel.}
    \label{fig:series1}
\end{figure}

The probe photon energy is naturally sampled between $\sim$512 and $\sim$518~eV due to the inherent shot-to-shot fluctuation of the self-amplified spontaneous emission (SASE) process. 
For each subset of XFEL shots at similar pump and probe photon and pulse energies, we identify the residual photoemission spectrum recorded at delays close to $\tau$~=~0 ($\tau$~=~-1~\textrm{to}~+1~fs), compared to a late delay reference ($\tau$~=~25~fs).
Each residual spectrum is normalized to the number of shots that contribute to it.
We integrate this residual photoemission spectrum across the full electron kinetic energy spectrum, aggregate the residual spectra across different pulse energies, and observe a sharp resonant feature within the bandwidth of the probe pulse, see panel \textbf{a} in \fref{fig:series1}.
We partition this feature into two regions (shaded red and blue), corresponding to the expected energetic position of the core-to-IVH transitions for the 6$a$ (red) and 7$a$ (blue) orbitals ($\sim$514.1 eV and $\sim$516.3 eV, respectively).
The gray shaded area demarcates an off-resonant region with good counting statistics which does not correspond to any expected resonance feature.
The error bars shown are the standard error across the independent pump and pulse energy bins, as outlined above.

We investigate the temporal dependence of the residual electron signal at the two defined positions shown in panel \textbf{a} of \fref{fig:series1}.
Panels \textbf{b} and \textbf{c} in \fref{fig:series1} show the percentage change in signal above 501~eV kinetic energy in the residual electron spectrum.
This corresponds to the region where the yield of normal Auger-Meitner electrons (measured during the experimental run at photon energy 650 eV) is highest, and the background photoelectrons are suppressed (see Fig. \ref{fig:ExperimentalLayout}).
We observe a time-dependent residual signal at the two identified resonance positions.
As in panel \textbf{a}, the error bars in both panels show twice the standard error across the independent subsets of XFEL shots defined by the binning procedure described above.
We fit the two traces to an exponentially modified Gaussian and enforce a common time of arrival, t$_0$, and x-ray pulse duration across both photon energies.
The extracted t$_0$ is --4.0 $\pm$ 1.3~fs and the extracted instrument temporal response (convolution of pump and probe pulse durations) is 6.3 $\pm$ 2.8~fs (FWHM) respectively.
The time axis is defined by the voltage applied to the dispersive magnetic chicane through which the electron bunch passes, and the extracted value of t$_0$ along this axis is within the expected bounds for the fresh slice mode of operation \cite{lutman2016fresh}.
The FWHM x-ray pulse duration is consistent with the nominal x-ray pulse pump and probe duration of 7~fs for these measurements.
The decay time extracted from the fit at the low energy side of the resonance (panel \textbf{c}) is 5.1 $\pm$ 3.2 fs. 
The decay time at the high energy side of the resonance position, 13.4 $\pm$ 7.0 fs, is significantly longer.

The normalized yield of Auger-Meitner electrons is roughly twice as large in panel \textbf{c} as it is in panel \textbf{b}.
This is consistent with the dipole moments calculated for transitions from the oxygen $1s$ orbital to the 6$a$ and 7$a$ orbitals within the ADC(2)x framework.
The calculated magnitude of the dipole moment for transitions to the 6$a$ orbital is roughly twice as large as for the 7$a$ orbital.
The relatively large error bars on the experimental fit are a result of the low counting statistics in the data and the limited spectral resolution afforded by the natural SASE bandwidth of the probe pulse (see Fig. \ref{fig:ExperimentalLayout}).

\begin{figure}[h]
    \centering
    \includegraphics[]{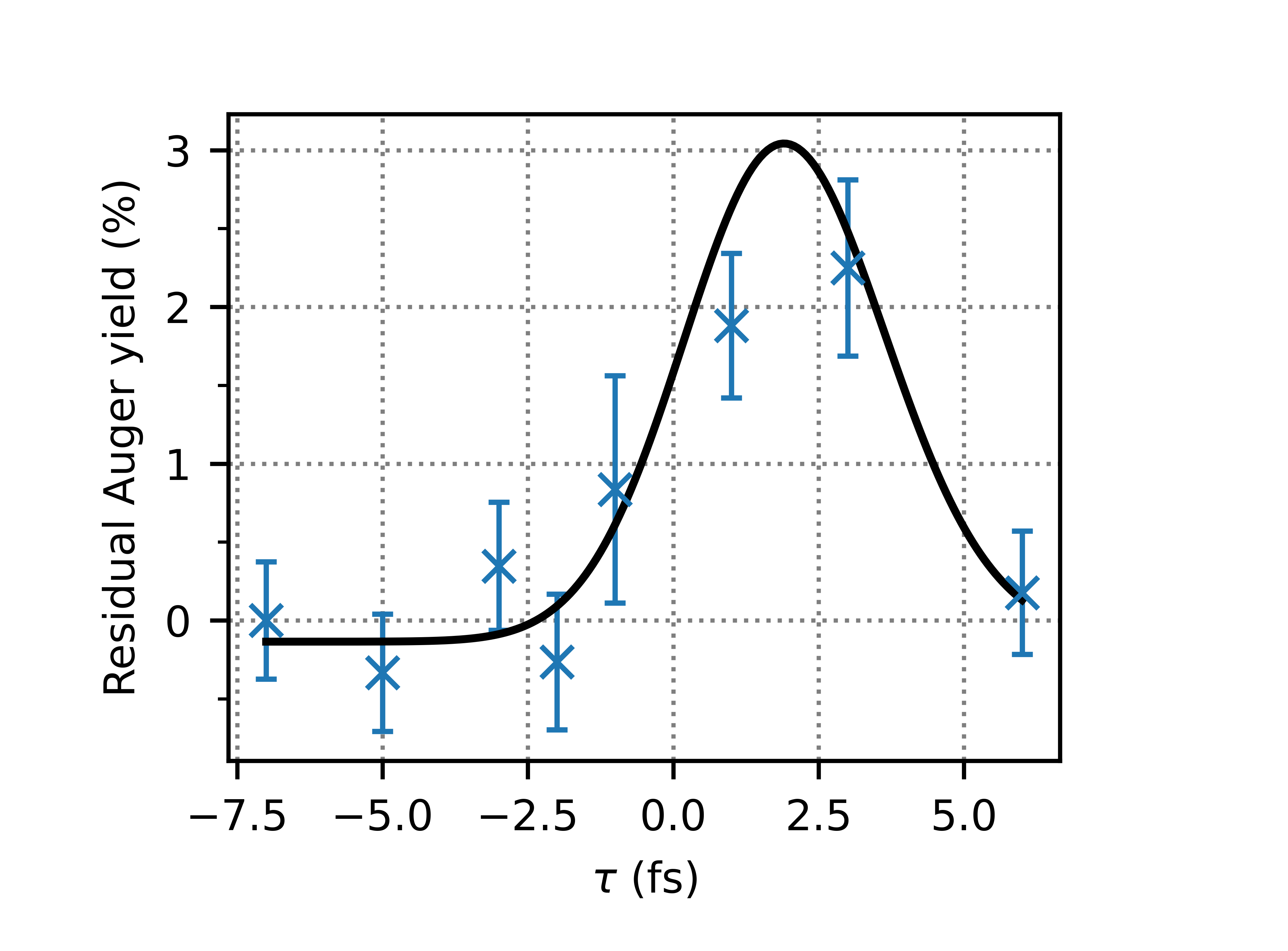}
    \caption{Time-dependence of excess Auger-Meitner electrons for a resonant probe, compared to a non-resonant probe. The signal is integrated over the region of peak flux, between binding energies of 17 and \SI{25}{\eV}, revealing a time dependent resonant Auger-Meitner signal. A fitted exponential decay convolved with a Gaussian measurement response function is fitted by bootstrapping across time and energy bins (see Methods) and is shown in black.
    This resonance, which maps the 6$a$ hole population, reveals a lifetime limited by the measurement response function of 1 $\pm$ \SI{2}{\femto\second}.}
    \label{fig:LCLSSignal}
\end{figure}

The lifetime extracted for the 6$a$ hole is close to the instrument response limit. 
For future measurements, identifying non-exponential decay dynamics will require pulses shorter than the lifetime of the target state. In an additional data set we use shorter (\SI{2.5}{\femto\second}) x-ray pulses \cite{lutman2018high} to measure the short lived 6$a$ orbital resonance with increased timing resolution. The lower statistics in this dataset precluded the same analysis as for the results in Fig. \ref{fig:series1}. The difference between the on-resonance signal with a central probe photon energy determined to fall in the range 513 to \SI{515}{\eV} with a similar number of shots lying outside this range was used to determine the excess electron spectra time-dependence plotted in \fref{fig:LCLSSignal}. The most prominent feature with a time dependent signal above the background lies within the electron binding energy energy range from 17 to \SI{25}{\eV}, consistent with where the Auger-Meitner spectrum has a maximum. 

The fractional increase in Auger-Meitner electron yield with binding energies in this range is shown in \fref{fig:LCLSSignal}. The measured rise time is significantly faster than that measured in Fig. \ref{fig:series1}, consistent with shorter x-ray pulses. From an exponential fit to the integrated signal, we obtain a decay constant of $1 \pm 2$ fs and a Gaussian instrument response function of 3.8 $\pm $ \SI{1.6}{\fs} full-width at half-maximum, where the errors denote one standard deviation in fit parameters estimated via bootstrapping (see Methods).
This instrument response function is consistent with the estimated parameters of \SI{2.5}{\femto\second} pulses.

The differential Auger-Meitner electron yield measurements require a large number of x-ray shots to gather enough data to establish statistically significant signals. 
As a consequence of long integration times there are a limited number of data points in the scan. 
The statistical error in the residual Auger-Meitner yield above the direct photoelectron background and the small number of delays that could be measured in the time available accounts for the uncertainties in the extracted value of the decay.
Nevertheless, these measurements demonstrate that with the shortest x-ray pulses available to us we can measure the population of the transient states associated with impulsive ionization, utilizing the inherent jitter of our probe to isolate the transient population of different initial hole states.

\section{Dynamics of inner valence electron hole states in isopropanol}\label{sec:theory}
We simulate the underlying electron dynamics expected to be resolved from the XFEL measurements by treating the photoionization dynamics in isopropanol within the sudden approximation~\cite{suddenapproximation} using an \textit{ab initio} theoretical method algebraic diagrammatic construction (ADC(2)x) \cite{trofimov2005molecular}. The advantage of this methodology is the ability to capture dynamics that are the result of electron correlation effects, as the  wavefunctions of the cation can be expressed in terms of intermediate states of either 1-hole ($\phi_i$) or 2-hole-1-particle ($\phi_{ij}^a$) character:
\begin{align}\label{eq:adcwavefunction}
    |\Psi^{(N-1)}\rangle = \sum_i C_i\phi_i + \sum_{i<j,a}C_{ij}^a \phi_{ij}^a.
\end{align}
Here the $i,j$ index indicate occupied Hartree-Fock orbitals in the neutral i.e. hole indices in the cation, and $a$ virtual orbitals in the neutral, or particle indices in the cation.

The inclusion of $\phi_{ij}^a$ configurations in the cation are vital, not just for describing the orbital relaxation effects upon ionization, but also the correlation driven dynamics, particularly frustrated Auger-Meitner configurations (FACs)~\cite{Bagus2004}, that can be observed in these small molecules upon sudden ionization. 
For cationic states produced by the removal of electrons from inner valence orbital this results in strong configuration interaction between $\phi_i$ and $\phi_{ij}^a$ configurations, leading to what is often termed the breakdown of the molecular orbital picture of ionization \cite{CederbaumMObreakdown} - the states produced are no longer well characterized by a particular 1-hole configuration in the ionized molecule.

For a short ionizing pulse with sufficiently high photon energy, the outgoing electron leaves the molecular volume on the few tens of attoseconds time scale.
The resulting cationic state is then expressed within the sudden approximaton as:
\begin{align}
    |\Psi^{(N-1)}_{\textrm{sudden}}\rangle = & \hat a_k|\Psi_0^{(N)} \rangle = \sum_n c_n |\Psi_n^{(N-1)}\rangle, \\
    c_n = & \langle \Psi_n^{(N-1)} |\hat a_k|\Psi_0^{(N)}\rangle.
\end{align}
Here, $\hat a_k$ is the annihilation operator associated with removing an electron from the $k^{th}$ molecular orbital of the neutral molecule, $\Psi_0^{(N)}$ is the ground state of the neutral, and $\Psi_n^{(N-1)}$ are the eigenstates of the cation. Within this approximation therefore, the system at time $t=0$ can be viewed as being in a superposition of cationic eigenstates. 
This system will evolve with time, and can be temporally characterized by the survival probability of the initial state $S(t)$:
\begin{align}
    S(t) =&\left|\left\langle\Psi^{(N-1)}_{\textrm{sudden}}(t=0)\left|\Psi^{(N-1)}_{\textrm{sudden}}(t)\right .\right\rangle\right|^2\nonumber\\=& \left |\sum_n c_n^2 \exp(-iE_nt) \right|^2
\end{align}
Full spatial and temporal characterization can be achieved by the calculation of time dependent hole densities~\cite{BreidbackTimeHoleDensity}, or the time dependent spin density~\cite{mendive2013coupled,VacherSpinDensity}.

Previous theoretical work has demonstrated the importance of zero point vibrational motion in modeling the time evolution of the hole dynamics~\cite{vacher2015electronPRA,vacher2017electron,Jenkins2016}.
Even for the case of dynamics arising from a superposition of just two electronic states, which should give rise to simple oscillations, because the molecule is in fact in low-lying vibrational states, the electronic coherences do not have a single, well-defined, frequency~\cite{vacher2015electronPRA}. 
This effect results in the significant degradation of the initial coherence~\cite{vacher2017electron}, and potentially decoheres any electronic dynamics faster than the nuclear motion subsequent to ionization. We accounted for this for the states of interest by sampling the dynamics for a large number of initial geometries over the range of the zero-point spread.

\begin{figure}[h]
    \centering
    \includegraphics[width=0.46\textwidth]{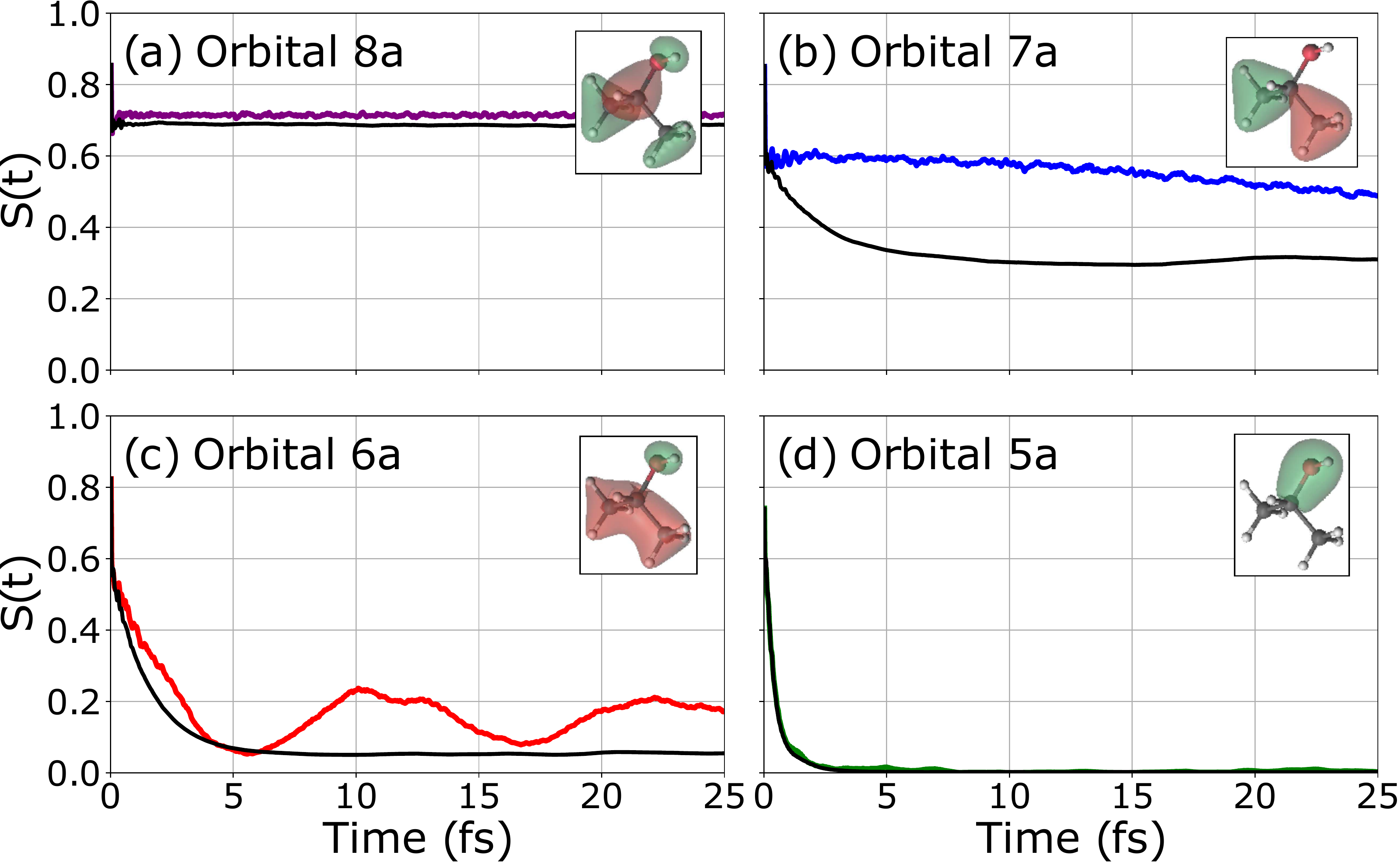}
    \caption{Hole survival probability of different hole states arising from removal of an electron from (neutral) inner valence orbitals 8$a$ (purple, outermost) to 5$a$ (green, innermost) after sudden ionization of isopropanol. The coloured survival probability curves correspond to a mixture (1:1) of the two lowest conformers of the molecule. The black curves correspond to the averaged dynamics over 500 geometries at a temperature of 300 Kelvin. The Hartree Fock orbitals of the neutral are shown in inserts for the gauche conformer.}
    \label{fig:HoleSurvivalProbability}
\end{figure}

For comparison, we also analyzed the dynamics expected upon removal of an electron from an inner valence orbital in the isopropanol molecule computed at the equilibrium geometry (gauche) and the anti conformer which lies 7.8 meV higher in energy \cite{kahnconformers}. Due to the conformers being energetically very similar, we considered a 1:1 ratio of the two geometries.  
We show these results in Figure \ref{fig:HoleSurvivalProbability}.

\section{Discussion}
The resulting dynamics expected from the removal of an electron from different initial inner valence orbitals upon ionization of isopropanol, can show very different behaviours.
\fref{fig:HoleSurvivalProbability}(a)~(purple curve) shows that the survival probability of the initial states produced by removing an electron from the 8$a$ orbital remains static for the two conformer averaged geometries, which can be interpreted as the hole remaining in the 8$a$ orbital. This is conserved when averaging over the zero-point sampling geometries (black curve).

In contrast, ionization from the more deeply bound orbitals, such as the 5$a$ orbital shown by the green curve in  \fref{fig:HoleSurvivalProbability}(d) (two conformer geometry average), which lie above the double ionization threshold, show very fast exponential decay resulting from Auger-Meitner transitions. The survival probability drops to zero which can be interpreted as the 5$a$ hole becoming completely filled again with an electron, consistent with the Auger-Meitner transition.
The 5$a$ orbital is filled by an outer valence electron and a secondary electron is emitted. Again these dynamics are conserved when averaging over the zero point sampling geometries (black curve).  

Between these two limiting cases, the temporal evolution resulting from removal of an electron from either the 6$a$ (\fref{fig:HoleSurvivalProbability}(c), red curve) or 7$a$ (\fref{fig:HoleSurvivalProbability}(b), blue curve) molecular orbitals shows more complicated dynamics. These are the two hole states probed in our measurements and for which a larger or smaller degree of breakdown of the molecular orbital picture of ionization is anticipated. For the case of removal of an electron from the 6$a$ orbital, the initially localized hole decays over a few femtoseconds followed by a shallow oscillatory evolution with a period of \SIrange{10}{15}{\fs} in the two-conformer geometry average. In contrast when an electron is removed from the 7$a$ orbital, the resulting superposition of states show a slow decay spanning tens of femtoseconds. Both the initial hole states created when removing an electron from these orbitals are below the double ionization threshold and therefore the Auger-Meitner relaxation channel is energetically forbidden. Instead what happens are frustrated Auger-Meitner transitions, where the inner valence hole is filled by an outer valence electron and a second electron is excited to a virtual orbital, but is still bound to the molecular volume. This is not a fully relaxed state and therefore there is a chance that the reverse transition happens and the hole reappears in the inner valence orbital where it started, leading to a partial revival of the initial state. By careful selection of the $\phi_{ij}^a$ configurations in the description of the cationic states, we have previously shown for trans-butadiene \cite{CooperAverbukhJCP2014} that this type of transition can account for the dynamics observed in the inner valence region. 

It should be noted that even in the two geometry sample there is a considerable difference in the dynamics expected from removal of an electron from the $7a$ orbital, as it is particularly sensitive to the breaking of the symmetry between gauche and anti conformers. This symmetry breaking is the rotation of the OH moiety. For the anti conformer the dynamics are similar to those seen in the case of removal of an electron from the $8a$ orbital; following the orbital relaxation, the superposition of states remains stationary with a high component of the hole remaining in the $7a$. The effect of breaking the symmetry in the gauche conformer leads to the slow oscillatory dynamics with a period of 10s of femtoseconds.

We include the zero point geometry spread by averaging over the distribution of nuclear geometries, which is expected to lead to a very fast dephasing on top of the few-fs electronic diffusion of the hole and thus to the predicted damping of the coherent revivals.
Removal of an electron from the 6$a$ orbital in the averaged geometry regime still shows a rapid decay with a similar lifetime compared to the two conformer average, but no partial revivals of the initial state. Although the hole survival probability strictly deviates from a simple exponential decay, for quantitative comparison we extract an effective hole lifetime (1/$e$ time constant, also referred to as decay time in this work) of 2.3 fs by means of a fit to the quasi-exponential behavior of the first five fs.
When an electron is removed from the $7a$ orbital, the averaged geometry picture shows a rapid but partial decay of the initial state, with this superposition remaining approximately static after 7 femtoseconds. This behavior is in complete contrast to the dynamics calculated at the equilibrium geometry which showed a very slow decay. 

The disturbance of the electronic charge density as a result of ionization will also induce nuclear dynamics in the molecule, which may also be very fast \cite{Baker,HJW,Mikosch}. This effect is not included in our calculations, as each of the sampled geometries uses the fixed-nuclei approximation when the survival probability is calculated. Nuclear motion after ionization will in reality modify all the survival probabilities, potentially modifying the quasi-exponential decay rate to help localize the hole on a timescale of 10s of femtoseconds.

Our measurements provide a confirmation of the behaviour found from the calculations for the $6a$ hole state, demonstrating that this is a transient hole state with only a few-femtosecond lifetime before relaxing into other states. We measure mutually consistent lifetimes of 5.1 $\pm$ 3.2 fs (Figure 3) and 1 $\pm$ 2 fs (Figure 4) in the two series of measurements with longer and shorter probe pulses respectively. The calculations suggest that this behaviour is dominantly driven by electron correlation dynamics and that the frustrated Auger-Meitner transitions are responsible for the very brief lifetime. Zero-point geometry effects will mask the partial revival dynamics in this case but the fast decay component is confirmed in the measurements. This is in contrast to another recent study in glycine where oscillatory dynamics were observed \cite{schwickert2020electronic}. The electron correlation mediated decay of the $6a$ hole state is too fast for post ionization nuclear motion to play a dominant role in the observed dynamics.
This is not necessarily the case for the 7a hole state, where a significant hole population survives zero-point energy effects. Indeed, for the 7a hole state the experimental measurements show a significantly longer decay ($\sim$13 fs).
This is in contrast to the theoretical simulations, based only on the geometry averaged electron dynamics, which predict an initially fast decay followed by a stabilization.
We therefore attribute this 13 fs decay to the missing 
relaxation channels in our simulation; channels opened up through post-ionization nuclear motion \cite{Marciniak}.
With our empirical data, which is limited in the number of statistics at each datapoint and the number of datapoints, it is not possible to separate the decay into different timescales; however, we postulate that decoherence from the spread of nuclear geometries dominates at the earliest timescales but that nuclear motion soon takes effect, such that a stabilization of hole population is not sustained.

To further investigate the contribution from the nuclear dynamics we performed auxiliary measurements using a VUV-IR measurement scheme (see Section VII). In this experiment, the time-dependence of CH$^{+}_{3}$ fragments with significant kinetic energy release was recorded. The time-dependent yield of these fragments exhibits exponential decay with a decay constant of 13 fs. Because the high-KER fragments arise from Coulomb explosion of dicationic molecules induced by the strong IR field, their time dependence probes the nuclear dynamics on the dissociating cationic potential energy surface by mapping the C-C bond distance to the KER of the fragments. The multiphoton nature of this fragmentation channel strongly favours the contribution of the states closest to the double ionization threshold, the 7$a$ and 6$a$ hole states. Although the 6$a$ hole state decays very rapidly according to our X-ray measurements, and so it is tempting to attribute the dominant contribution to the longer lived 7$a$ hole state, an alternative explanation is suggested by recent work ~\cite{Marciniak}. The frustrated Auger-Meitner decay of 6$a$ will result in the population of the longer-lived 2h1p satellite states, which will contribute to the high KER channel observed in our VUV-IR measurement in addition to contributions from 7$a$ 2h1p satellites.

Whilst both the IR and x-ray probe methods used in this work have the potential to yield a comparable temporal resolution in measuring the inner valence hole state evolution following photoionization, the former does not permit any information to be directly obtained on the instantaneous localization of the electronic hole. In contrast the use of an x-ray core-to-IVH transition in the probe step allows the extra information on the instantaneous localization of the hole state to be established in a direct way. In particular, utilizing the resonance in the dipole transitions, we have been able to separate dynamics associated with ionization of specific valence orbitals. In this way, we were for example able to disentangle the rapid relaxation of the states associated with ionization of the 6$a$ orbital, which occurs as a consequence of the set of correlation-driven satellite peaks.

Moreover, the IR probe requires a relatively strong (\SI{\sim 5e12}{\watt\per\square\centi\meter}) field to induce efficient ionization to the double-continuum, which may in itself perturb significantly the fragile electronic correlation dynamics due to being near-resonance with various valence shell transitions. In contrast, the x-ray probe is essentially weakly perturbative and far from resonant with any valence transitions. 

Similarly the x-ray probe, which does not drive vibrational or rotational transitions as a strong field IR probe would, preserves the spread of the nuclear geometries present in the ensemble. While this means that the longer lived oscillations, which may be present at a single molecular geometry level, are not observed, the rapid decay is reliant on the electron correlations which lead to the coherent superposition of eigenstates. This is not to say that all the hole states do not survive over the longer timescales; our IR-probe measurements demonstrate that the hole state (nominally 7$a$) relaxes over tens of femtoseconds.

All valence hole states are created by the pump pulse in both the x-ray pump-probe measurements and the auxiliary XUV-IR measurements. In the former the measurement specificity comes from the resonant tuning of the probe x-rays, that isolates the channel of interest, and this is central to the choice of this measurement scheme. We argue also that the multi-photon nature of the probe in our auxiliary measurement also leads to a discrimination in favour of the states closest to the double-ionization limit (dominated by 6$a$ and its satellites but with some contribution from the 7$a$ hole), but in this case the degree of specificity is less and cannot be readily controlled.
Our XUV/IR measurement indicates that nuclear dynamics following inner valence ionization of isopropanol occur on a timescale that is long enough to have a negligible effect on the very fast decay of the 6$a$ hole, consistent with the x-ray pump/x-ray probe measurements above.
The contribution to the high-KER CH$^{+}_{3}$ fragments is expected to come from 6$a$ and 7$a$ 2h1p satellite states. The decay lifetime identified by our XUV/IR measurements supports the hypothesis that the hole delocalization following ionization to these states is significantly affected by nuclear motion because the electronic timescales for these are longer than those for the 6$a$ hole state.
This nuclear dynamics contribution also offers an explanation for the deviation from theory for 7$a$ between our x-ray pump - x-ray probe measurement and our theoretical calculations including the inherent spread of initial geometries in our molecular ensemble but excluding post-ionization nuclear motion \cite{Marciniak}.

\section{Conclusion and Outlook}

Our measurements with an x-ray pump-probe methodology have observed a highly transient hole corresponding to the removal of an electron from inner valence orbitals in the ionization of isopropanol. We have been able to demonstrate the capability to distinguish between the dynamics arising from the removal of an electron from the 6$a$ and 7$a$ orbitals using the variation of the probe photon energy. The extracted decay times for the 6$a$ hole state survival probability from our two series of measurements with different x-ray pulse duration are 5.1$\pm$3.2 fs and $1 \pm 2$ fs.
Good agreement is obtained for the observed dynamics arising from removal of an electron from the 6$a$ orbital with simulations produced from considering zero-point energy spread; our theoretical calculations performed within the ADC(2)x framework predict 2.3 fs. 
This indicates that the nuclear motion post ionization is too slow to exert a significant effect on the dynamics of the  6$a$ hole state survival probability. In contrast, consideration of post-ionization nuclear motion is required to fully understand the dynamics observed when an electron is removed from the 7$a$ orbital.
In particular, the longer-lived hole population, which survives beyond the initial decay and is predicted by our calculations, is not observed in our measurement. This is posited to be a result of decoherence due to nuclear motion following inner valence ionization.

The timescale of C-C bond stretching following inner valence ionization is identified in our auxiliary XUV/IR experiment.
Unlike our new x-ray pump - x-ray probe technique, the XUV/IR experiment does not allow for specificity of the ionized inner valence orbital.
By consideration of the order of photon processes required to produce the CH$^{+}_{3}$ fragments, we anticipate this measurement to predominantly consist of cationic states produced from the removal of an electron from the 6$a$ orbital.
Our x-ray pump-probe measurements for the $6a$ hole state present the first observation of the temporal signature of frustrated Auger-Meitner transitions. The complex nature of the frustrated Auger-Meitner decay process that results in the population of the 2h1p satellites of the 6$a$ and 7$a$ orbitals may offer the possibility that these longer lived states, not directly probed in our state specific x-ray pump-probe measurement, contribute to the KER signal with a similar timescale nuclear motion to that observed in the x-ray pump-probe measurement of the 7$a$ hole state. .

We stress that 6$a$ and 7$a$ hole states dynamics are both affected by coupling to the nuclear modes; however, in the former, electronic dynamics are so fast that nuclear motion is not the dominant cause of the observed dynamics - and instead the correlated electron dynamics and the zero-point geometry spread dephasing are playing the main role in what is observed. In contrast 7$a$ hole state has significantly slower electronic dynamics (after the weak initial coherence decay) and here nuclear dynamics are expected to play a strong role - and we find the evidence for this in our x-ray pump-probe measurements.

The same methodology of x-ray pump-x-ray resonant probing can be applied with the newly developed enhanced SASE attosecond mode \cite{zholents2004proposal} (not available at the time of our experiment) recently demonstrated to generate pulses of 480 attosecond duration in this photon energy range \cite{duris2020tunable}. This will allow the full resolution of dynamics anticipated from the theoretical calculation for this hole state as well as the other inner valence states 5$a$, 7$a$ and 8$a$ that span the behaviour from fast Auger-Meitner decay to Koopmans' \cite{koopmans1934zuordnung} like steady state behaviour.
As the bandwidth in this case may exceed the spacing between these states, the possibility of exciting a superposition of states from distinct orbital groups also arises.

Although in the case of the isopropanol molecule the hole dynamics are not migratory in the sense that the hole moves from one well localized atomic center to another, they are migratory in the sense that the well localized hole becomes delocalized through the frustrated Auger-Meitner transitions. The same method will be suitable for systems where the electronic wave packet moves through the molecular structure (case (a) in the discussion of Section I) and can track the hole migration through the molecule in time and space at one or even multiple atomic sites. The potential to probe charge migration and subsequent photochemical activity using x-ray pump-probe methodology with XFEL generated sub-femtosecond pulses will deepen our understanding of the first moments following the photo-ionization of matter. This understanding is crucial to other fields of science such as photo-chemistry, photo-biology and atmospheric science.

\begin{acknowledgments}
Use of the Linac Coherent Light Source (LCLS), SLAC National Accelerator Laboratory, is supported by the U.S. Department of Energy, Office of Science, Office of Basic Energy Sciences under contract No. DE-AC02-76SF00515. The authors acknowledge the invaluable support of the technical and scientific staff of the LCLS at SLAC National Accelerator Laboratory. JPM, TB, OA, BC, TD, DG, AS-G, LJF, VA, MR, MB,MV, JWGT, DW, SJ, CB, EWL, ASJ, PM-H, DW acknowledge funding from EPSRC through grants EP/I032517/1, EP/N018680/1 and EP/R019509/1, and ERC Advanced Grant ASTEX (2012-17). AA, GD, YK, and CB acknowledge support from the U.S. Department of Energy, Office of Basic Energy Sciences, Division of Chemical Sciences, Geosciences, and Biosciences through Argonne National Laboratory. Argonne is a U.S. Department of Energy laboratory managed by UChicago Argonne, LLC, under contract DE-AC02-06CH11357. TJAW, JPC, and NB were supported by the U.S. Department of Energy Office of Science, Basic Energy Science Division.  NB under grant DE-SC0012376 (UConn). LK, JER, MA, CS gratefully acknowledge support by the Swedish Research Council (VR) under grants No. 2014-04518 and 2018-04088. RF and RJS thank the Swedish Research Council and the Knut and Alice Wallenberg Foundation, Sweden for financial support. PK acknowledges financial support by the Czech Science Foundation (Project GAČR No. 17-10866S).KU acknowledges support  via the X-ray Free Electron Laser Utilization Research Project and the X-ray Free Electron Laser Priority Strategy Program of MEXT, 
the Dynamic Alliance for Open Innovation Bridging Human, Environment and Materials Program,
and the IMRAM Program of Tohoku University. 

\end{acknowledgments}

\section{Methods}

\subsection{Experimental Methods}

The main data-sets, presented in a Section II, both use an x-ray pump and x-ray probe and were taken at the LCLS free electron laser employing the two pulse/two color fresh slice operating mode \cite{lutman2016fresh} focused at the AMO endstation. These two data-sets differ in their nominal pulse durations and the range of delays covered. The data presented in Figure \ref{fig:LCLSSignal} is taken with shorter pulse durations (approximately \SI{2.5}{\femto\second}) over a shorter delay range whilst the data presented in Figure \ref{fig:series1} uses longer pulses (\SI{7}{\femto\second}) and covers larger delays.

The x-rays were generated with an electron beam energy of \SI{3.96}{\giga\eV} and the undulators and chicane operated in the fresh slice two color mode with a chicane adjustable delay from between \SI{-15}{\fs} to \SI{+25}{\fs}. This provided the two pulses, one (the pump) around \SI{502}{\eV}, the other the probe at around the target resonances at \SI{514}{\eV} or \SI{516}{\eV}. 

Pump and probe spectra were recorded for every shot using a downstream x-ray spectrometer and used to establish the central photon energies, bandwidth, and relative pulse energies of the pump and probe pulses.
The spectrometer was an adapted version of the instrument described in \cite{nordgren1989soft}, with a 1200 l/mm grating operated in the Rowland geometry imaging a mechanical slit onto a microchannel plate coupled to a phosphor screen.
A gas-detector was used to extract the absolute pulse energies. The correlation between the x-ray spectrometer and the gas-detector confirmed that there was good x-ray pointing stability to ensure the x-ray spectrometer measurements were a high-fidelity diagnostic of the relative pulse intensities. The photon spectrum had an intrinsic spread and varies from shot-to-shot over a band of \SI{\sim5}{\eV} which allowed for the post-selection of shots based on photon energy and was used to distinguish between the resonances (for the longer pulse data series) and with off-resonant shots for comparison.

An x-ray transverse RF cavity streaking method (X band transverse deflecting cavity; XTCAV) \cite{ratner2015time} diagnostic could be used to monitor the pulse duration and jitter from the nominal delay settings, although for the shorter pulse series it had insufficient temporal resolution and so was used primarily to determine that the jitter was comparable to the pulse durations in the longer pulse series data. Simulation for parameters of this series indicates that the pulses are approximately \SI{2.5}{\fs} in duration \cite{guo2020simulation}. The observed temporal jitter from the XTCAV analysis was $\pm$\SI{2}{\fs} for the short pulse data series, permitting us to use the chicane delay setting as the primary tool for timing the delay.

Both pulses were focused to a spot size of \SI{\sim 2}{\micro\meter} using a series of Kirkpatrick-Baez mirrors. The isopropanol was prepared at 298 K and introduced into the focus of the x-rays with a \SI{50}{\micro\meter} diameter gas needle with careful monitoring of the constancy of the partial pressure through the experiment. The sample was delivered into the interaction region in the form of a low-density effusive gas beam. The photo- and Auger-Meitner electrons were measured using a hemispherical electron analyzer (Scienta) with an energy resolution \SI{\sim 0.1}{\eV} over the range of electron energies \SIrange{470}{510}{\eV} (the region of interest) and correction made over this range for instrument response. Due to the need to identify a relatively weak signal (O 1$a$ Auger-Meitner electrons) against a large background of photoemission from many other channels it required to use a large number of laser shots at each delay to accumulate adequate measurement statistics (see Data Analysis). We therefore address a limited number of delays, between -6 and 7 fs in approximately 2 fs steps in the shorter pulse data-set and between -15 and \SI{25}{\femto\second} in 13 steps concentrated at close to zero delay in the longer pulse data-set.
To analyze the data the shots were selected to fall within a restricted range of pump-probe pulse energy ratios, the data normalized accounting for other variations (see Data Analysis).

\subsection{Data Analysis}

Free electron laser data are intrinsically noisy due to the SASE emission's dependence on the quantum stochastic nature of spontaneous emission and fluctuations in the properties of the electron injection into the undulators, leading to different pulse parameters (energies, intensities, spectral width, etc.).
Shot-to-shot fluctuations lead to noisy signals, thus requiring increased statistics; and slower changes, such as small drifts which occur when tuning to change the time delay, can lead to time dependent artifacts. 
To negate possible artifacts caused by variation in parameters across different shots and different delay points, we sort the measured photoelectron spectra according to the single-shot pulse energy and photon energy of the x-ray pulses.

The data-set taken with longer x-ray pulses (Fig. \ref{fig:series1}) has sufficiently high counting statistics to allow a four-dimensional binning of the XFEL shots according to the measured pump pulse energy, probe pulse energy, time delay and central photon energy of the probe pulse.
Because the central photon energy of both pulses is well correlated in fresh slice operation (we measure $\rho$~=~0.75 in this experiment), this effectively represents a binning in pump central photon energy as well.
To avoid artifacts induced by the natural fluctuations inherent to XFEL operation, we determine the time-dependent residual photoelectron spectrum for each different pump pulse energy, probe pulse energy, and probe central photon energy. 
We use a time late ($\tau$~=~25 fs) measurement as reference.
The probe photon energy-dependence of the yield of this time-dependent residual signal at times close to $t_0$ ($\tau$~=--1 fs, 1 fs) is investigated and found to show a sharp peak within the probe bandwidth.
The expected transition energies from the oxygen 1$a$ to the 6$a$ and 7$a$ orbitals lie within this peak. 
To isolate signal due to a resonant probe interaction only, we compare the residual spectrum measured at probe photon energies on-resonance with expected core-to-IVH transitions with the residual spectrum measured at off-resonant probe photon energies.
We monitor the high kinetic-energy ($>$~501 eV) photoelectron yield because this energetic region corresponds to the peak of the normal Auger-Meitner spectrum (measured at 650~eV incident photon energy in the same experiment), and has very limited contamination from other photoionization processes arising from the pump and probe pulses.
We identify the time-dependence of this yield across the full temporal range sampled by the x-ray pump-x-ray probe experiment.

For the dataset with shorter x-ray pulses (Fig. \ref{fig:LCLSSignal}), the counting statistics were much lower.
To account for this, we applied a variety of normalisation and binning techniques.
The effectiveness of this depends on the quality of the diagnostics, and can be measured by looking at the variance of the signal.
For the normalization, this means finding the functional dependence of the signal, i.e. the electron kinetic energy spectra, on the diagnosed parameters. 
The two most important parameters to correct for were the total pulse energy, measured using a low density N$_2$ gas detector \cite{Hau-Riege2008}, and the pump-probe intensity ratio, measured from the ratio of signal over appropriate spectral regions of the x-ray spectrometer.
Given that the two-photon signal was expected to be a small feature on a dominant single-photon (photo-electron) signal, a linear normalization scheme was chosen for the pulse energy normalization.
A large negative time delay was used for comparison. The reduced counting statistics of this data-set gave need to consider Auger-Meitner electrons at lower kinetic energies (489 -- 497 eV), where the Auger-Meitner yield is largest but the direct photoelectron background is also more significant.
For this reason, variation in the pump-probe intensity ratio required more sophisticated normalization as the dependence of the electron spectra on the pump-probe signal varies across the binding energy range of the electrons. Simply put, parts of the photoelectron spectrum dominated by pump excited electrons will increase with pump-probe ratio whilst probe dominated regions will decrease. 
Therefore, a linear dependence was found individually for each energy bin, resulting in a `spectrum' of scaling factors and offsets. This works well provided that the dependence is indeed linear, which is a good approximation in the case when shots are limited to a band of pump-probe ratios by the shot discrimination.  

Despite the normalization and the randomization of the order of the run delays, there was still concern that variation of the FEL parameters between runs could have led to artifacts in the data.
The pump probe ratio was of particular concern.
The simplest method of reducing this variation is to only allow shots within a specified, narrow band of values.
This was done for the gas detector and the XTCAV delay.
Doing this is effective, albeit some delay dependence persists if there is a skewed distribution within this narrow band, and there is an expensive reduction in the statistics.
Therefore, an improved method was used:
Firstly, a desired pump-probe ratio was chosen.
Then shots outside of a broader band centred on this value were discarded.
Remaining shots were binned according to their pump-probe ratio, then the bin bounding the desired value was identified. 
The minimum of the counts in the bin to the left of this bin and the counts in the bin to the right of this bin was then found and excess shots discarded from the more numerous bin. This was repeated for the bins 2 to the left/right, 3 to the left/right, etc.
In this way, a symmetric distribution of shots centred around a desired value was reached, which was constant across all pump-probe delays.

Since the signal is independent of Auger-Meitner electron kinetic energy, the error bars shown in figure were estimated from the standard deviation across electron kinetic bins. The error in the exponential fit parameters were estimated by bootstrapping the electron kinetic energy vs time delay data set. Here the 928 data points were resampled (with replacement) 5000 times and an exponential fit was performed on each sample data set. The median and standard deviation of the 5000 fits is then computed to give the results presented in the main text.

\subsection{Theoretical Methods}

We modelled the average dynamics that one might expect given the uncertainty in the initial geometry at the point of ionization by the probe pulse. The Molpro Quantum Chemistry package~\cite{werner2012molpro} was used to optimize the geometry of isopropanol and calculate the normal mode vibrational frequencies, at the Hartree-Fock level and using cc-pVTZ basis set. The Newton-X package \cite{barbatti2014newton,barbatti2007fly} was then used to generate 500 sample geometries at a temperature of 300 K. In this model the coordinates and momenta are sampled according to their probability distribution assuming a harmonic vibrational state. For the vibrational ground state these initial conditions should match the Wigner distribution for the quantum harmonic oscillator.

We treat each of the sampled geometries independently, and re-calculate for each the survival probabilities shown in \fref{fig:HoleSurvivalProbability}. We use the same ADC(2)x description of the wavefunctions of the cationic states (Eq.~\ref{eq:adcwavefunction}), assume sudden ionization, and again use the frozen-nuclei approximation during the time propagation, neglecting the effects of nuclear motion after the ionization event.

\section{Auxiliary measurement: strong-field probing of inner valence hole dynamics}
\label{sec:VUV-IR}

The auxiliary measurement utilised a VUV pump and NIR probe, was conducted using a carrier-envelope phase stabilised titanium doped sapphire based chirped pulse amplifier (Femtolaser) of \SI{4}{\femto\second}, centred at \SI{750}{\nano\meter}, to form the IR probe field and also to drive a gas based high harmonics generation (HHG) source to generate the the VUV pump \cite{fabris2015synchronized,frank2012invited}. The spectral bandwidth necessary to generate these few cycle pulses was achieved by propagation in a gas filled hollow core fibre and subsequent compression. The HHG spectrum had a  \SI{30}{eV} cutoff and was filtered from its driving laser with an aluminium foil, filtering energies below \SI{17}{eV}. 
The intensity of the NIR probe field was set below the threshold for multiphoton ionization of neutral isopropanol molecules, i.e. corresponding to an intensity around $5 \times 10^{12}$ Wcm$^{-2}$. 

Previously it has been proposed that multiphoton laser enabled Auger-Meitner decay (LEAD)~\cite{LEAD2011Ranitovic} or single photon spLEAD~\cite{spLEAD2013Cooper} schemes could be used to measure the hole dynamics that occur in molecules upon ionization of inner valence electrons, by tracing the survival probability onto the observation of secondary Auger-Meitner electrons. 
Recently, sp-LEAD transitions have been observed in atoms \cite{observedspleadPrince}. Here we use a few-cycle laser pulse to induce Coulomb explosion fragmentation of cations formed by XUV pulses by excitation to the dication. This provides information on the instantaneous C-C bond length in the cation, and hence nuclear dynamics, but does not directly probe the electron hole.

We ionized isopropanol molecules with attosecond~($\sim650$~as) vacuum ultraviolet~(VUV) pulses generated via HHG.  
The spectrum of the VUV pulse extended from \SI{17}{eV} to \SI{30}{eV}, which is sufficiently energetic to remove an electron from the 5a orbital and those less tightly bound.
Such a large coherent bandwidth is sufficient to excite a superposition of states, corresponding to an initial configuration of $5a$, 6$a$, or $7a$, but not a superposition between these configurations.
The XUV pump will ionize all of the valence states, and in principle the dynamics for the ensemble will be correspondingly complex, however here again there is a high degree of probe specificity. The broad band IR probe causes multi-photon ionization for the $6a$ and 7a states the ionization to the double continuum will involve 3 and 4 photons respectively, in contrast for the 8a state we require 5 photons and for 9a 6 photons. 
Given our intensity, we are in the perturbative regime, so the signal will be strongly dominated by direct multiphoton ionization of 6a, 7a as well as satellite 2h1p like channels that are generated via the lowest order process.
Contributions from 8a and other states higher in energy are suppressed by the higher orders of multiphoton ionization required.

The VUV pulses were measured by streaking to have a duration of 650 as \cite{barillot2017towards}. The pump and probe were co-focused using a two-part piezo activated mirror into an isopropanol gas jet. The inter-pulse delay could be controlled with a few tens of attosecond precision using the piezo controlled motion of a split mirror system. The NIR probe pulse will further ionize any cation states created by the VUV pulse that are still surviving when it arrives, to produce dicationic final products via one and two-photon excitation. Detection of the charged fragments carrying significant kinetic energy from Coulomb explosion is used as the detection channel of the resulting dications.
The VUV-IR delay was controlled with \SI{10}{as} precision in a range of $-20$ to $+60$ fs.
Fragment ions were collected using a Wiley-McLaren~\cite{WM} ion time-of-flight spectrometer. 
The spectrometer has sufficient momentum resolution to observe the kinetic energy release (forward-backward going ions giving rise to satellites around charge/mass peaks) associated with any fragment. 
\begin{figure}[]
    \centering
    \includegraphics[]{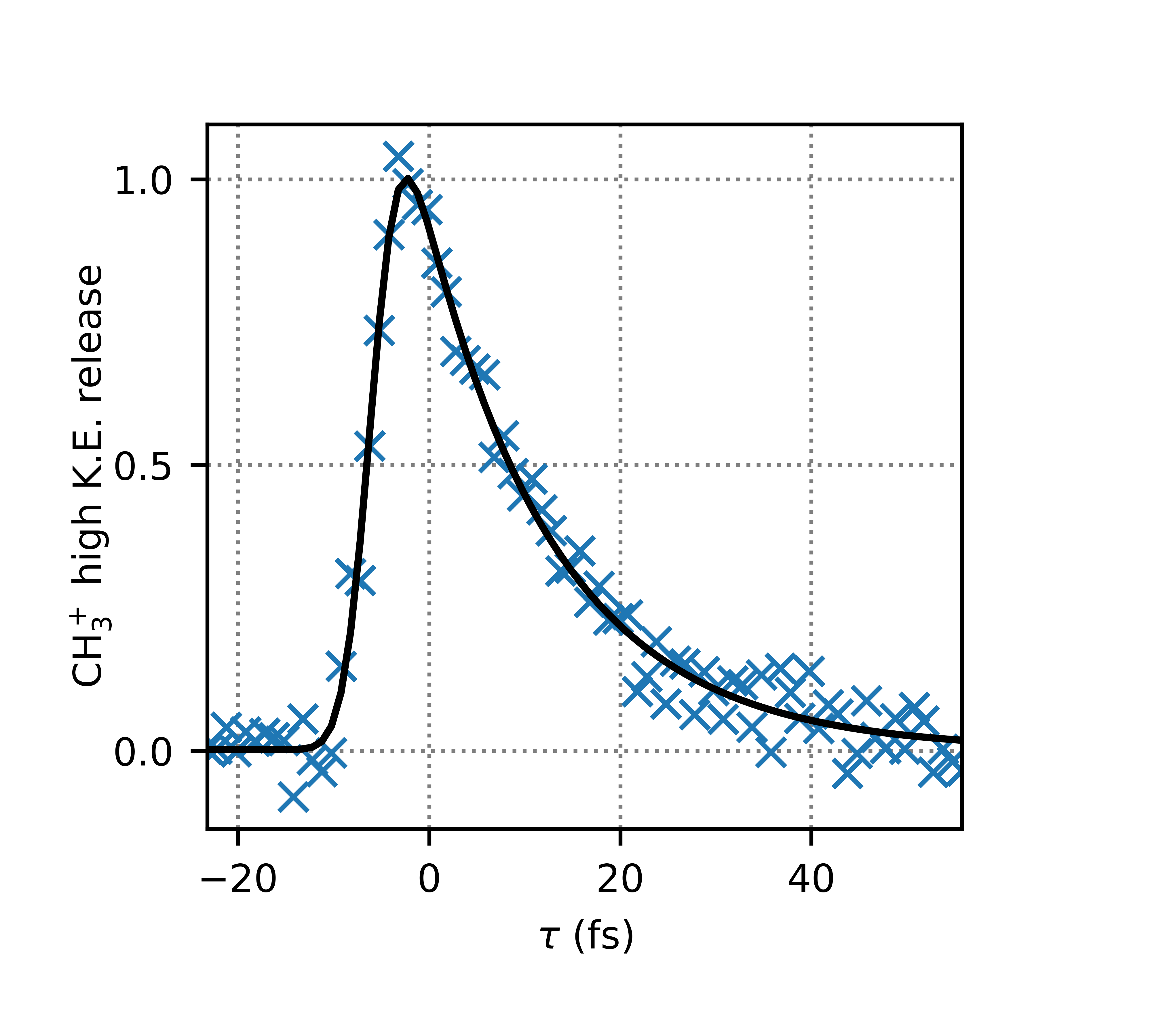}
    \caption{Transient high kinetic energy CH$_3^+$ ion yield, attributed to the decay of the intermediate cationic channel in isopropanol using a VUV pump at \SI{20}{\eV} and 600 attosecond duration and a \SI{4}{\fs} IR probe to induce transitions to the second ionization continuum}
    \label{fig:ImperialSignal}
\end{figure}

We found a fast transient signal in the yield of CH$_3^+$ fragment ions with a high kinetic energy release, as shown in Figure \ref{fig:ImperialSignal}.
The extracted decay time of this transient is \SI{13}{\fs}. 
The high kinetic energy release is consistent with a Coulomb explosion of the dication species, and implies that this fragmentation channel arises from multiphoton dissociative ionization by the intense IR probe.
Since this proceeds via the intermediate cation state we can infer that nuclear dynamics are proceeding in this state so as to increase the distance between the partner ions at the moment of the Coloumb explosion and so quench the yield of high kinetic energy release fragments.
Due to difficulties in accurately modeling the strong-field ionization of the cationic states, from this measurement it is difficult to confirm that there is no perturbation of the electronic dynamics by the probe. 

\bibliography{Isopropanol,referencesJPC}

\end{document}